\documentclass[12pt]{article}
\usepackage{amssymb,amsmath}
\usepackage{txfonts}
\usepackage{amsfonts}
\usepackage{latexsym}
\usepackage{graphicx,color,subfigure}
\usepackage[english]{babel}

\newcommand{\be}{\begin{equation}}
\newcommand{\ee}{\end{equation}}
\newcommand{\ba}{\begin{array}}
\newcommand{\ea}{\end{array}}
\newcommand{\bqa}{\begin{eqnarray}}
\newcommand{\eqa}{\end{eqnarray}}

\renewcommand{\d}{\mathrm{d}}

\def\baselinestretch{1.4}
\def \be {\begin{equation}}
\def \ee {\end{equation}}
\def \bea {\begin{eqnarray}}
\def \eea {\end{eqnarray}}
\def \nn {\nonumber}

\def \a {\alpha}
\def \b {\beta}

\def \G {\Gamma}
\def \d {\delta}
\def \eps {\epsilon}
\def \m {\mu}
\def \n {\nu}
\def \k {\kappa}

\def \s {\sigma}
\def \r {\rho}
\def \o {\omega}

\def \th {\theta}
\def \Th {\Theta}

\def \t {\tau}
\def \dag {\dagger}
\def \p {\partial}

\def\bd{
\begin{document}}
\def\ed{\end{document}}
\def\nn{\nonumber}
\def\bea{\begin{eqnarray}}
\def\eea{\end{eqnarray}}
\let\bm=\bibitem
\let\la=\label

\def\N{{\cal N}}
\def\sst{\scriptscriptstyle}
\def\thetabar{\bar\theta}
\def\Tr{{\rm Tr}}
\def\one{\mbox{1 \kern-.59em {\rm l}}}

%
%%%%%%%%%%%%%%%%%%%%%%%%%%%%%%%%%%%%%%%%%%%%%%%%%%%%%%%%
%%                       Abbreviations for Greek letters

\def\a{\alpha}      \def\da{{\dot\alpha}}
\def\b{\beta}       \def\db{{\dot\beta}}
\def\c{\gamma}  \def\C{\Gamma}  \def\cdt{\dot\gamma}
\def\d{\delta}  \def\D{\Delta}  \def\ddt{\dot\delta}
\def\e{\epsilon}        \def\vare{\varepsilon}
\def\f{\phi}    \def\F{\Phi}    \def\vvf{\f}
\def\h{\eta}
\def\k{\kappa}
\def\l{\lambda} \def\L{\Lambda}
\def\m{\mu} \def\n{\nu}
\def\o{\omega}
\def\bo {\bar{\o}}
\def\P{\Pi}
\def\r{\rho}
\def\s{\sigma}  \def\S{\Sigma}
\def\t{\tau}
\def\th{\theta} \def\Th{\Theta} \def\vth{\vartheta}
\def\X{\Xeta}
\def\z{\zeta}
\def\w{\wedge}
\def\u{\underline}
\def\hs{\hspace}
\def\G{\Gamma}

%%%%%%%%%%%%%%%%%%%%%%%%%%%%%%%%%%%%%%%%%%%%
%%                      Calligraphic letters

\def\cA{{\cal A}} \def\cB{{\cal B}} \def\cC{{\cal C}}
\def\cD{{\cal D}} \def\cE{{\cal E}} \def\cF{{\cal F}}
\def\cG{{\cal G}} \def\cH{{\cal H}} \def\cI{{\cal I}}
\def\cJ{{\cal J}} \def\cK{{\cal K}} \def\cL{{\cal L}}
\def\cM{{\cal M}} \def\cN{{\cal N}} \def\cO{{\cal O}}
\def\cP{{\cal P}} \def\cQ{{\cal Q}} \def\cR{{\cal R}}
\def\cS{{\cal S}} \def\cT{{\cal T}} \def\cU{{\cal U}}
\def\cV{{\cal V}} \def\cW{{\cal W}} \def\cX{{\cal X}}
\def\cY{{\cal Y}} \def\cZ{{\cal Z}}

%%%%%%%%%%%%%%%%%%%%%%%%%%%%%%%%%%%%%%%%%%%%
%%                    Underline letters

\def\ua{\underline{\alpha}} \def\ubb{\underline{\beta}}
\def\ug{\underline{\gamma}}
\def\ub{\underline{\phantom{\alpha}}\!\!\!\beta}
\def\uc{\underline{\phantom{\alpha}}\!\!\!\gamma}
\def\um{\underline{\mu}} \def\un{\underline{\nu}}
\def\ud{\underline\delta}
\def\ue{\underline\epsilon}
\def\una{\underline a}\def\unA{\underline A}
\def\unb{\underline b}\def\unB{\underline B}
\def\unc{\underline c}\def\unC{\underline C}
\def\und{\underline d}\def\unD{\underline D}
\def\une{\underline e}\def\unE{\underline E}
\def\unf{\underline{\phantom{e}}\!\!\!\! f}\def\unF{\underline F}
\def\unm{\underline m}\def\unM{\underline M}
\def\unn{\underline n}\def\unN{\underline N}
\def\unp{\underline{\phantom{a}}\!\!\! p}
\def\unP{\underline P}
\def\unq{\underline{\phantom{a}}\!\!\! q}
\def\unQ{\underline{\phantom{A}}\!\!\!\! Q}
\def\unH{\underline{H}}
\def\ul{\underline}
%%%%%%%%%%%%%%%%%%%%%%%%%%%%%%%%%%%%%%%%%%%%
%%                  Slash letters

\def\As {{A \hspace{-6.4pt} \slash}\;}
\def\bs {{b \hspace{-6.4pt} \slash}\;}
\def\Ds {{D \hspace{-6.4pt} \slash}\;}
\def\ds {{\del \hspace{-6.4pt} \slash}\;}
\def\ss {{\s \hspace{-6.4pt} \slash}\;}
\def\ks {{ k \hspace{-6.4pt} \slash}\;}
\def\ps {{p \hspace{-6.4pt} \slash}\;}
\def\pas {{{p_1} \hspace{-6.4pt} \slash}\;}
\def\pbs {{{p_2} \hspace{-6.4pt} \slash}\;}

%%%%%%%%%%%%%%%%%%%%%%%%%%%%%%%%%%%%%%%%%%%%
%%              hatted letters

\def\Fh{\hat{F}}
\def\Vh{\hat{V}}
\def\Xh{\hat{X}}
\def\ah{\hat{a}}
\def\xh{\hat{x}}
\def\yh{\hat{y}}
\def\ph{\hat{p}}
\def\xih{\hat{\xi}}

%%%%%%%%%%%%%%%%%%%%%%%%%%%%%%%%%%%%%%%%%%%%
%%          tilde letters
\def\psit{\tilde{\psi}}
\def\Psit{\tilde{\Psi}}
\def\tht{\tilde{\th}}

\def\At{\tilde{A}}
\def\Qt{\tilde{Q}}
\def\Rt{\tilde{R}}
\def\Nt{\tilde{N}}

\def\at{\tilde{a}}
\def\st{\tilde{s}}
\def\ft{\tilde{f}}
\def\pt{\tilde{p}}
\def\qt{\tilde{q}}
\def\vt{\tilde{v}}
\def\nt{\tilde{n}}

%%%%%%%%%%%%%%%%%%%%%%%%%%%%%%%%%%%%%%%%%%%%%%%%%%%%%%%%%%
%%          bar             %%

\def\delb{\bar{\partial}}
\def\bz{\bar{z}}
\def\bD{\bar{D}}
\def\bB{\bar{B}}

%%%%%%%%%%%%%%%%%%%%%%%%%%%%%%%%%%%%%%%%%%%%%%%%%%%%%%%%%%
%%          bold                %%

\def\bk{{\bf k}}
\def\bl{{\bf l}}
\def\bp{{\bf p}}
\def\bq{{\bf q}}
\def\br{{\bf r}}
\def\bx{{\bf x}}
\def\by{{\bf y}}
\def\bR{{\bf R}}
\def\bV{{\bf V}}

%%%%%%%%%%%%%%%%%%%%%%%%%%%%%%%%%%%%%%%%%%%%%%%%%%%%%%%%%%
%%                      Miscellaneous                   %%
%%%%%%%%%%%%%%%%%%%%%%%%%%%%%%%%%%%%%%%%%%%%%%%%%%%%%%%%%%

\def\d{\delta}\def\D{\Delta}\def\ddt{\dot\delta}

\def\p{\partial} \def\del{\partial}
\def\xx{\times}
\def\uno{\mbox{1 \kern-.59em {\rm l}}}

\def\trp{^{\top}}
\def\inv{^{-1}}
\def\dag{{^{\dagger}}}
\def\pr{\prime}
\textwidth=6.0in \hoffset=-.3in \textheight=9in \voffset=-.8in
\def\baselinestretch{1.2}

\def\rar{\rightarrow}
\def\lar{\leftarrow}
\def\lrar{\leftrightarrow}

%%%%%%%%%%%%%%%%%%%%%%%%%%%%%%%%%%%%%%%%%%%%
%% Document
%%%%%%%%%%%%%%%%%%%%%%%%%%%%%%%%%%%%%%%%%%%%%

\title{ On Self-Dual Warped AdS$_3$  Black Holes }
\author{Bin Chen and Bo Ning\footnote{Email: bchen01, ningbo@pku.edu.cn}\\ \\
{\small Department of Physics} \\
{\small and State Key Laboratory of Nuclear Physics and Technology}\\
{\small Peking University, Beijing 100871, P.R.China}}
\date{}
\bd \maketitle
\begin{abstract}
We study a new class of solutions of three-dimensional topological
massive gravity. These solutions can be taken as non-extremal black
holes, with their extremal counterparts being discrete quotients of
spacelike warped AdS$_3$ along the $U(1)_L$ isometry. We study the
thermodynamics of these black holes  and show that the first law is
satisfied. We also show that for consistent boundary conditions, the
asymptotic symmetry generators form only one copy of the Virasoro
algebra with central charge $c_L = \frac{4\nu\ell}{G(\nu^2+3)}$,
with which the Cardy formula reproduces the black hole entropy. We
compute the real-time correlators of scalar perturbations and find a
perfect match with the dual CFT predictions. Our study provides a
novel example of warped AdS/CFT correspondence: the self-dual warped
AdS$_3$ black hole is dual to a CFT with nonvanishing left central
charge. Moreover our investigation suggests that the quantum
topological massive gravity asymptotic  to the same spacelike warped
AdS$_3$ in different consistent ways may be dual to different 2D
CFTs.

\end{abstract}

\newpage
%\begin{document}

\section{Introduction}

Three-dimensional topological massive gravity (TMG)\cite{Deser:1981wh,Deser:1982vy} is described by the following action
\be
I_{TMG} = \frac{1}{16\pi G}\left[\int d^3x\sqrt{-g}\,(R+2/\ell^2)+{\frac{1}{\mu}}I_{CS}\right]~,
\ee
where $I_{CS}$ is the gravitational Chern-Simons action and we take both $G$ and
$\mu$ positive. For every value of the coupling $\mu$, TMG admits an AdS$_3$ vacuum solution of radius $\ell$, which
is known to be perturbatively unstable except at the chiral point $\mu\ell=1$ \cite{Li:2008dq}. However, for generic
values of the coupling $\mu$, it has been suggested that the theory could possess other stable backgrounds, namely the
spacelike, timelike or null warped AdS$_3$ spaces \cite{Andy08}. The spacelike warped
AdS$_3$ admits the left-broken isometry group $U(1)_L\times SL(2,\mathbb{R})_R$, and there exist black hole solutions
which can be obtained by performing discrete identifications in this background. For $\mu\ell>3$, the spacelike warped AdS background is said to be stretched and the black holes are regular. It has been conjectured in \cite{Andy08}
that TMG defined with suitable ``warped AdS'' boundary conditions would be dual to a two-dimensional CFT with central
charges
\be
  c_L = \frac{4 {\nu} \ell}{G( {\nu}^2 + 3)}\,, \qquad c_R = \frac{(5 {\nu}^2 + 3)\ell}{G {\nu}({\nu}^2 + 3)}
\ee
where $\nu=\mu\ell/3$. It was shown that these central extensions can be derived from the asymptotic symmetry algebra
associated to the spacelike warped AdS geometries \cite{Compere:2008cv,Compere:2009zj,Blagojevic:2009ek}, giving support
to the warped AdS/CFT correspondence. Further support to this conjecture comes from the study of the quasi-normal modes and real-time correlators
of the spacelike warped AdS$_3$ black hole in \cite{ChenXu09,ChenXu2,{Chen:2009cg}}. One subtle point in this correspondence
is that the asymptotic geometry of the warped black hole is related to the one of the global warped AdS$_3$ by a local coordinate transformation,
which then induces the identification of quantum numbers to set up the dictionary of the warped AdS/CFT correspondence. Similar thing
happens in the null warped case.
%With certain consistent boundary conditions, the algebra of asymptotic symmetries is isomorphic to the semi-direct sum
%of a Virasoro algebra and an algebra of currents.

\vspace{3mm}

In this paper we provide another significant support to the conjectured warped AdS/CFT correspondence by studying a new
kind of \emph{self-dual warped AdS$_3$} black hole. Their extremal counterparts have been exhibited in different coordinates
in \cite{Andy08}, where an heuristic expression of their entropy is given in the form of Cardy's formula, suggesting the
existence of a dual CFT. Our self-dual warped black hole metric takes a similar form to the near-NHEK geometry
\cite{Bredberg:2009pv} and is asymptotic to the spacelike warped AdS$_3$ without any coordinate transformation. We show that for appropriate boundary conditions,
the asymptotic symmetry generators form one copy of the Virasoro algebra with central charge $c_L$\,, with which the
application of the Cardy formula \cite{Cardy:1986ie} precisely reproduces the Bekenstein-Hawking entropy. This result is in favor of the
warped AdS/CFT conjecture, in a chiral version though.

\vspace{3mm}

The motivation of proposing the self-dual warped black hole solution comes from the Kerr/CFT correspondence \cite{Guica:2008mu}, which could be taken as a generalization of the warped AdS/CFT correspondence. Geometrically, a slice of NHEK geometry at fixed polar angle is locally a self-dual warped AdS$_3$ spacetime. It was suggested that the quantum gravity in NHEK is dual to a CFT with only left-moving temperature. Moreover, further study suggests that the near-horizon geometry of the near-extremal Kerr black hole (Near-NHEK) is dual to the same CFT with nonvanishing right temperature, as the right-moving sector gets excited~\cite{Bredberg:2009pv}. The near-NHEK geometry differs slightly from the NHEK geometry in the AdS$_2$ sector. We are inspired to ask whether
there are  near-NHEK like solutions in TMG. It turns out that such metrics indeed solve the equation of motion of TMG, and their extremal counterparts
are precisely the self-dual solutions found in \cite{Andy08}. These solutions are locally equivalent to the spacelike warped AdS$_3$ by a singular coordinate transformation. This is in accordance with the fact that all the solutions of TMG are locally AdS or warped AdS if they are asymptotically related\footnote{all Einstein solutions in three dimensions are locally AdS. We are unaware of a theorem generalizing this to locally warped AdS for suitable asymptotic behavour.}.

\vspace{3mm}

Strictly speaking, our solutions do not belong to the category of regular black holes, which require the existence of a geometric or causal singularity shielded by an event horizon. In fact, the solutions are free of curvature singularity and regular everywhere, even though there exist Killing horizons corresponding to the Killing vector of time translations. The situation is reminiscent of AdS$_2$ black hole.
On the other hand, we show that similar to the black holes,  these spacetimes have the regular thermodynamic behavior, satisfying the first law of thermodynamics. We find that the conserved charges and the entropy is independent of the right-moving sector, suggesting the dual CFT is chiral. However we also notice that in the dual CFT both left-moving and right-moving temperatures are nonvanishing. The right-moving one can be explained in a similar way as the temperature of the AdS$_2$ black hole \cite{Spradlin:1999bn}, while the left-moving one originates from the identification of the angular coordinate.

\vspace{3mm}

 Similar to the extreme Kerr case, the asymptotic symmetry group is an enhancement of the $U(1)_L$ isometry. However, there exists significant difference from the Kerr case on consistent asymptotic boundary conditions. Actually since the context here is TMG rather than Einstein gravity, we need to propose a set of slightly different boundary conditions to account for the correction from the Chern-Simons term. More interestingly, we find that the asymptotic symmetry generators form only one copy of the Virasoro algebra with nonvanishing central charge. This is quite different from the case of the spacelike warped AdS$_3$ black hole, studied in \cite{Compere:2008cv,Compere:2009zj,Blagojevic:2009ek}, where both left and right central charges are nonzero, even though the left central charge in both cases are the same.  Obviously our study provides a different playground to explore
the warped AdS/CFT correspondence.

\vspace{3mm}

In the next section we introduce the self-dual warped AdS$_3$ black holes in TMG. In section 3 we study their black hole
thermodynamics and verify that the first law holds. In section 4 we yield the left and right temperatures of the dual CFT
and explain their different origin. We specify the boundary conditions in section 5, show that the centrally
extended asymptotic symmetry algebra gives a left central charge $c_L = \frac{4\nu\ell}{G(\nu^2+3)}$, with which
the application of the Cardy formula reproduces the Bekenstein-Hawking entropy. We further test the warped AdS/CFT
correspondence in section 6 by calculating the scalar real-time correlator from gravity, finding perfect agreement with the
CFT prediction. We end with discussions in section 7.

\section{Self-dual warped black hole}

\subsection{Topologically massive gravity}

The action of three-dimensional topologically massive gravity with a negative cosmological constant is
\be
I ~=~ \frac{1}{16\pi G}\int d^3x\sqrt{-g}\left(R+2/\ell^2\right)
 ~+~\frac{\ell}{96\pi G \nu}\int d^3x\sqrt{-g}\varepsilon^
{\lambda\mu\nu}\Gamma^\rho_{\lambda \sigma}\left(\partial_{\mu}\Gamma^
\sigma_{\rho \nu}+\frac{2}{3}\Gamma^\sigma_{\mu\tau}\Gamma^\tau_{\nu
\rho} \right)\,,\label{Action}
\ee
where $\varepsilon^{\tau \sigma u}=+1/\sqrt{-g}$ is the Levi-Civita
tensor and $G$ has positive sign. The dimensionless coupling $\nu$ in the coefficient
of the Chern-Simons action is related to the graviton mass $\mu$ by
\be \nu=\frac{\mu \ell}{3} \,.\ee
Without loss of generality we take $\nu$ to be positive. Varying the above action with respect to the metric
gives the equation of motion
\be
{G}_{\mu\nu}-\frac{1}{\ell^2}g_{\mu\nu}+{\frac{\ell}{3\nu}}C_{\mu\nu}=0\,,
\label{EOM}
\ee
where $G_{\mu \nu}$ is the Einstein tensor and $C_{\mu \nu}$ is
  the Cotton tensor:
\be
C_{\mu\nu} = \varepsilon_\mu^{~\alpha \beta}\nabla_\alpha\left(R_
{\beta\nu}-\frac{1}{4}g_{\beta\nu}R\right)\,.
\ee
Any Einstein solution with $G_{\mu \nu}=g_{\mu\nu}/
\ell^2$ is also a solution of TMG; there are also non-Einstein solutions that satisfy
the equation of motion.

\subsection{Self-dual warped black hole solution}

An interesting class of non-Einstein black hole solutions of TMG, which are asymptotic to spacelike warped AdS$_3$ spacetime,
is the following
\begin{multline}\label{BH}
ds^2 ~=~ \frac{\ell^2}{\nu^2 + 3}\left(-(x-x_+)(x-x_-)\,d\tau^2 + \frac{1}{(x-x_+)(x-x_-)} \,dx^2\right.
\\\left.+\frac{4 \nu^2}{\nu^2 +3} ( \alpha d\phi + (x-\frac{x_+ + x_-}{2})\, d\tau)^2\right)\,,
\end{multline}
where the coordinates range as $\tau\in[-\infty,\infty]$, ${x}\in[-\infty,\infty]$ and $\phi \sim \phi + 2\pi$. In the metric, $\alpha$ is a free parameter, which will be seen to be related to
the entropy and the left-temperature of the black hole.  The Killing vectors are manifestly the spacelike $\partial_\phi$ and $\partial_\tau$. A coordinate shift in $x$ and reciprocal rescaling of $x$ and $\tau$ brings the metric to a canonical form $x_+=-x_-=1$, the difference $x_+-x_-$ is thus seen to correspond to a choice of time $\tau$. We will refer to these as
\emph{self-dual warped} black holes, in analogy to the self-dual solutions studied in \cite{Coussaert:1994tu}.
The horizons are located at $x_+$ and $x_-$ where $1/g_{xx}$ vanishes.
The vacuum solution for the black holes is given by $x_+ = x_- = 0$ and $\alpha=1$, which is
spacelike warped AdS$_3$ in Poincar\'e coordinates and under a periodic identification of $u=u+2\pi$. For $\nu^2>1$, these solutions are free of naked CTCs.
%In the appendix \ref{app:Kruskal} we extend the space beyond the horizon.

The self-dual warped black hole (\ref{BH}) is closely related to the
self-dual solution mentioned in \cite{Andy08}. The self-dual
solution in \cite{Andy08} is the discrete quotient of the spacelike
warped AdS$_3$ by identifying along the J$_2$ isometry. In
Poincar\'e-like coordinates, the metric is of the form
\begin{equation}\label{eq:SDPoinc}
 ds^2 = \frac{\ell^2}{\nu^2+3}\left( -\tilde{x}^2 d\tilde{\tau}^2 + \frac{d\tilde{x}^2}{{\tilde{x}}^2} + \frac{4\nu^2}{\nu^2+3}\left(\alpha\, d\tilde{\phi} + \tilde{x}d\tilde{\tau}\right)^2\right)~,
\end{equation}
with $\tilde{\phi}\sim\tilde{\phi}+2\pi$. The parameter $\alpha$ just characterizes a family of different quotients. The self-dual warped black
hole (\ref{BH}) is related to this metric through coordinate
transformation
\begin{eqnarray}
 \tilde \tau^\pm &\equiv & \tilde \tau \pm \frac{1}{\tilde
 x}=\tanh\left[\frac{1}{4}\left((x_+-x_-)\tau \pm \ln
 \frac{x-x_+}{x-x_-}\right)\right],\nonumber \\
 \tilde \phi &=& \phi +\frac{1}{2}\ln \left[\frac{1-(\tilde
 \tau^+)^2}{1-(\tilde
 \tau^-)^2}\right]. \label{coordtrans}
\end{eqnarray}
Globally, the maximal analytic extension of the self-dual warped
black hole is diffeomorphic to (\ref{eq:SDPoinc}). However, the
above coordinate transformations are singular at the boundary $x\to
\infty$, which indicates that the physics behind the two solutions
(\ref{eq:SDPoinc}) and (\ref{BH}) are different. The situation here
is very similar to the relation between near horizon geometry of
extreme Kerr (NHEK) and near horizon geometry of near-extreme Kerr
(near-NHEK)\cite{Bredberg:2009pv}, or the relation
between AdS$_2$ and AdS$_2$ black hole\cite{Spradlin:1999bn}. %Let us recall the definition of an AdS$_2$ black
%hole from \cite{Spradlin:1999bn}: ``an AdS$_2$ geometry with a prefered
%choice of time''. This choice, denoted $\tau$ in our paper, is what
%restricts the metric to the form where the lapse function has two single roots.
%as opposed to say Poincar\'e coordinates with a double root or global coordinates with no real roots.
In a sense, the self-dual warped AdS black hole may be taken as a
$U(1)$-fibred AdS$_2$ black hole. As in near-NHEK or AdS$_2$ black
holes, we will see that an observer in (\ref{BH}) at fixed $x$
measure a Hawking temperature proportional to $x_+-x_-$. Moreover,
as in near-Kerr case, the entropy does not depend on $x_+-x_-$, but
the scattering amplitudes do depend on $x_+-x_-$.

\subsection{Diffeomorphism}

%We note that the above self-dual black hole metric
%is locally related to the spacelike stretched AdS$_3$ black hole\cite{Andy08} via the linear coordinate
%transformation
%\bea\label{tr1}
%t&=&\frac{1}{\nu^2 + 3} \left(2\nu\alpha\phi - \left( (r_+ + r_-)\nu - \sqrt{r_+ r_- (\nu^2 + 3 )}\right) \frac{x_+ - x_-}{r_+ - r_-}\,\tau\right)\,,\nn\\
%r&=&\frac{r_+ - r_-}{x_+ - x_-}\left(x - \frac{x_+ + x_-}{2}\right) + \frac{r_+ + r_-}{2}\,,\nn\\
%\theta&=&\frac{2}{\nu^2 + 3} \frac{x_+ - x_-}{r_+ - r_-}\, \tau\,,
%\eea
%which interchanges the roles of time and angle.

The above self-dual warped black hole metric is locally equivalent to spacelike warped AdS$_3$
\be\label{wAdS}
{ds^2} = \frac{\ell^2}{\nu^2 + 3}\left(  - \cosh^2\sigma d\upsilon^2 + d
\sigma^2 + \frac{4\nu^2}{\nu^2 + 3} \left(du + \sinh\sigma d\upsilon
\right)^2 \right)
\ee
through the coordinate transformation
\bea\label{tr2}
\upsilon&=&\tan^{-1}\left[\frac{2\sqrt{(x-x_+)(x-x_-)}}{2x - x_+ - x_-}\sinh\left(\,\frac{x_+ - x_-}{2}\,\tau\right)\,\right]\,,\nn\\
\sigma&=&\sinh^{-1}\left[\frac{2\sqrt{(x-x_+)(x-x_-)}}{x_+ - x_-}\cosh\left(\,\frac{x_+ - x_-}{2}\,\tau\right)\,\right]\,,\nn\\
u&=&\alpha\phi\,+\,\tanh^{-1}\left[\frac{2x - x_+ - x_-}{x_+ - x_-}\coth\left(\,\frac{x_+ - x_-}{2}\,\tau\right)\,\right]\,.
\eea
In the coordinates of \eqref{wAdS}, the Killing vectors of spacelike warped AdS$_3$ are
\label{section:appA}
\begin{eqnarray*}
J_2 &=& 2\partial_{u}
\\
\tilde{J}_1 &=&  {2\sin{\upsilon} \tanh{\sigma} \partial_{\upsilon} - 2\cos{\upsilon} \partial_{\sigma}+\frac{2\sin{\upsilon}}{\cosh{\sigma}} \partial_{u} }  \\
\tilde{J}_2 &=&  {-2\cos{\upsilon} \tanh{\sigma} \partial_{\upsilon}-2\sin{\upsilon} \partial_{\sigma} - \frac{2\cos{\upsilon}}{\cosh{\sigma}} \partial_{u}}  \\
\tilde{J}_0 &=&  2\partial_{\upsilon}
\end{eqnarray*}
and satisfy the algebra $[\tilde{J}_1,\tilde{J}_2]=2\tilde{J}_0$, $[\tilde{J}_0,\tilde{J}_1]=-2\tilde{J}_2$,  $[\tilde{J}_0,\tilde{J}_2]=2\tilde{J}_1$ and $[J_2,\tilde{J}_{1,2,3}]=0$.

\vspace{3mm}

\section{Thermodynamics}

In this section we look into the thermodynamics of self-dual warped black holes. We will see that after accounting for
the effects of the Chern-Simons term, the various thermodynamic quantities obey the first law.

\subsection{Conserved charges}

The ADT mass $\mathcal{M}^{ADT}$ and angular momentum $\mathcal{J}^{ADT}$ of self-dual warped black hole can be calculated
following the procedure developed in \cite{Bouchareb:2007yx}, where they computed the conserved charges associated to the
Killing vectors $\partial_\tau$ and $\partial_\phi$ for a TMG solution linearized around an arbitrary background using the
surface integral expressions derived in \cite{Abbott:1982jh,Deser:2002rt,Deser:2002jk,Deser:2003vh}. The final result is
\bea\label{charge}
\mathcal{M}^{ADT} = 0\,,\quad\quad  \mathcal{J}^{ADT} = \frac{(\alpha^2 - 1)\nu\ell}{6G(\nu^2 + 3)}\,.
\eea
That is, the ADT mass and angular momentum do not depend on the parameters $x_+$ and $x_-$\,. The situation here is similar
to the case of near-NHEK geometry\cite{Bredberg:2009pv} of which the ADM mass defined in the asymptotically flat region is
given by its extremal value in that limit and does not depend on $T_R$\,. We will see that the entropy $S$ does not depend on
$x_+$ and $x_-$ either. However, the observers at fixed $x$ in self-dual warped black hole measure a Hawking temperature proportional
to $(x_+ - x_-)$\,.% similar to the case of near-NHEK geometry as well.

\subsection{Entropy}

The entropy of the self-dual warped black hole is composed of two terms, one comes from the Einstein action and the other is
due to the Chern-Simons contribution\cite{Solodukhin:2005ah,Tachikawa:2006sz}. The total entropy is given by
\be
S=S_E + S_{CS}=\frac{2\pi\alpha\nu\ell}{3G(\nu^2+3)}\,.
\ee

\subsection{First law}

It is a nontrival check of black hole thermodynamics that the mass and angular momentum as well as the entropy are related by the
first law. From the self-dual warped black hole metric (\ref{BH}) which is already in ADM form, we read the Hawking
temperature $T_H$ and the angular velocity of the horizon $\Omega_h$\,,
\be
T_H = \frac{x_+ - x_-}{4\pi\ell}\,,\quad\quad \Omega_h = - \frac{x_+ - x_-}{2\alpha\ell}\,.
\ee
Then we check explicitly that the first law
\be
d\mathcal{M}^{ADT}=T_H\, dS + \Omega_h \,d\mathcal{J}^{ADT}\,,
\ee
is satisfied for a variation of the black hole parameter \,$\alpha$.

\subsection{On the vacuum solution}

In the above discussion, we have taken the vacuum solution to be the one with $\a_0=1$.
However, we have no good reason to prefer this value. In fact, the computation of the entropy is independent of the choice of $\a_0$, while the computations on the ADT mass and angular momentum depends on the choice of the vacuum. Nevertheless, even with a different choice of $\a_0$, the first law of thermodynamics still holds.

From the entropy and the left temperature $T_L$ of the dual CFT, it seems that the natural choice of the vacuum solution should be $\a_0=0$. Unfortunately, in this case, the metric becomes degenerate. It would be nice to have a criterion to decide the vacuum solution.

\section{Temperatures}

In this section we will derive the left- and right-moving temperatures of the self-dual warped black holes and show that they have
different origins.

\vspace{3mm}

Since the metric (\ref{BH}) looks much similar to the near-NHEK geometry\cite{Bredberg:2009pv}, it is sensible to
define a quantum vacuum in analogy to the Frolov-Thorne vacuum\cite{Frolov:1989jh,Ottewill:2000qh,Duffy:2005mz}. The construction
begins by expanding the quantum fields in eigenmodes of the asymptotic energy $\omega$ and angular momentum $k$. Consider a scalar
field $\Phi$, we may write
\be
\Phi= \sum_{\omega, k,l} \phi_{\omega kl} e^{(-i\omega \tau+i k \phi)}\cR_l(x)\,.
\ee
After tracing over the region inside the horizon, the vacuum is a diagonal density matrix in the energy-angular momentum eigenbasis with a Boltzmann
weighting factor
\be\label{ggf}
{\rm exp}\left(-\hbar \frac{\omega-k \Omega_{H}}{T_H} \right).
\ee
This reduces to the Hartle-Hawking vacuum in the non-rotating $\Omega_H=0$ case. The left and right charges $n_L\,, \,n_R$ associated to $\partial_\phi$ and
$\partial_\tau$ are
\be n_L\equiv k\,,\quad\quad\quad n_R\equiv\omega\,. \ee
Identifying
\be {\rm exp}\left(-\hbar{\frac{\omega-k \Omega_H}{T_H}} \right)={\rm exp}\left(-{\frac{n_L}{T_L}}-{\frac{n_R}{T_R}}\right) \ee
defines the left and right temperatures
\be T_L={\frac{\alpha}{2\pi\ell}}\,,\quad\quad\quad T_R={\frac{x_+ - x_-}{4\pi\ell}}\,. \ee
In the extremal limit $ x_+ = x_-$, these reduce to
\be T_L={\frac{\alpha}{2\pi\ell}}\,,\quad\quad\quad T_R=0\,. \ee

\vspace{3mm}

The right temperature denotes the deviation from extremality. As discussed in detail in \cite{Maldacena:1998uz,Spradlin:1999bn} in
the AdS$_2$ context, the observers moving along worldlines of fixed $x$ will detect a thermal bath of particles at temperature $T_R$
in the global Hartle-Hawking vacuum, essentially similar to the case that Rindler observers detect radiations in the Minkowski vacuum. \nocite{Clement:2004yr,Moussa:2008sj}

\vspace{3mm}

The left temperature, on the other hand, arises from the periodical identification of points in warped AdS$_3$. From (\ref{tr2}) we see that
in order for the coordinate transformation to reproduce the black hole we need to identify points along the $\partial_\phi$ direction such
that $\phi\sim\phi + 2\pi$\,. Expressing the $\partial_\phi$ Killing vector in terms of the original warped AdS$_3$ coordinates, we discretely
quotient along the isometry
\be
\partial_\phi = \frac{\alpha}{2} J_2 = \pi\ell\, T_L J_2\,,
\ee
where  $J_2\in U(1)_L$ is the Killing vector.  In analogy with the BTZ
case\cite{Banados:1992gq,Maldacena:1998bw}, the coefficient of the shift is the temperature of the dual 2D CFT. Note that this periodical identification
makes no contribution to the right temperature.

\section{Asymptotic behavior}

Since the self-dual warped black hole solution is locally isometric to the spacelike warped AdS$_3$, it is natural to
expect the existence of a dual 2-dimensional CFT. In this section we will show that with suitable boundary conditions,
the asymptotic symmetry generators form one copy of a Virasoro algebra with central charge $c_L = \frac{4\nu\ell}{G(\nu^2+3)}$, which
is precisely what Anninos et al\cite{Andy08} conjectured for $\nu>1$ quantum TMG. An application of the Cardy
formula to the CFT density of states reproduces the black hole entropy.

\subsection{Boundary conditions}

We impose the following boundary conditions
\bea\label{BC}
\left(\begin{array}{ccc}
~~h_{\tau\tau}=O(1)&  \quad\quad h_{\tau x}=O(1/x^3) & \quad  h_{\tau\phi}=O(x\,) \\
h_{x\tau}=h_{\tau x}&  \quad\quad h_{x x}=O(1/x^3) & \quad~~~~h_{x \phi}=O(1/x\,)\\
h_{\phi\tau}=h_{\tau\phi} &  h_{\phi x}=h_{x \phi}  & \quad h_{\phi\phi}=O(1) \end{array} \right)
\eea
where $h_{\mu\nu}$ is the deviation of the full metric from the vacuum. Here the allowed deviations
$h_{\tau \phi}$ and $h_{\phi\phi}$ are of the same order as the leading terms in (\ref{BH}). In this regard, the
boundary conditions differ from both the spacelike warped AdS$_3$ boundary conditions \cite{Compere:2008cv,Compere:2009zj,Blagojevic:2009ek},
where all deviations are subleading, and from the Kerr boundary conditions \cite{Guica:2008mu}, where $h_{\tau\tau}$
rather than $h_{\tau \phi}$ is of leading order. The most general infinitesimal diffeomorphism which preserves the boundary
conditions (\ref{BC}) is of the form
\bea\label{asymkilling}
\zeta&=&[C + O(1/x^3)]\,\partial_{\tau} \,+\, O(1)\,\partial_x \,+\, [\epsilon\,(\phi) + O(1/x^2)]\,\partial_{\phi}\,,
\eea
where $\epsilon\,(\phi)$ is an arbitrary smooth function of the coordinate $\phi$ , and $C$ is an arbitrary constant. The subleading
terms indicated above can be seen to correspond to trivial diffeomorphisms by computing the generators. Therefore the asymptotic
symmetry group contains one copy of the conformal group of the circle generated by
\bea\label{ASG}
\xi_{\epsilon}&=&\epsilon\,(\phi) \,\partial_{\phi}\,.
\eea
This Virasoro algebra contains a $U(1)$ isometry subgroup. Since $\phi\sim\phi + 2\pi$, it is convenient to define
$\epsilon_n(\phi)=e^{i n \phi}$ and $\xi_n = \xi(\epsilon_n)$. These generators admit the following commutators
\bea\label{virasoro}
i\,[\xi_m,\,\xi_n]~=~(m-n)\,\xi_{m+n}\,,
\eea
and $\xi_0$ generates the $U(1)$ rotational isometry.

\vspace{3mm}

The allowed symmetry transformations (\ref{asymkilling}) also include $\tau$ translations generated by $\partial_{\tau}$\,.
The conjugate conserved charge $E_R$, with the expression given in the next subsection, could be well-defined by imposing
the supplementary boundary condition
\bea\label{suppBC}
2 \,\alpha\,h^{~~[0]}_{\tau\phi} - h^{~~[0]}_{\phi\phi}=0\,,
\eea
where $h^{~~[0]}_{\mu\nu}$ is the coefficient of the leading order deviation $h_{\mu\nu}$ in the neighboorhood of our metrics.

\vspace{3mm}

The boundary conditions (\ref{BC})(\ref{suppBC}) are very different from the usual spacelike warped AdS$_3$ boundary conditions \cite{Compere:2008cv,Compere:2009zj,Blagojevic:2009ek}. In the latter case, consistent boundary conditions were imposed in which
the $SL(2,\mathbb{R})$ isometry is enhanced to a Virasoro algebra and the $U(1)$ isometry is enhanced to a current algebra. In the
present situation, the $SL(2,\mathbb{R})$ isometry becomes trivial and the $U(1)$ isometry is enhanced to a Virasoro, similarly to
the case of Kerr \cite{Guica:2008mu}.
%We will see in the next subsection that the centrally extended Virasoro algebra gives the left central charge $c_L = \frac{4\nu\ell}{G(\nu^2+3)}$.

\subsection{Asymptotically conserved charges}

The conserved charge associated with an asymptotic Killing vector \,$\xi$\, is constructed in the covariant formalism by Barnich, Brandt and Comp\`{e}re
\cite{Barnich:2001jy,Barnich:2003xg,Barnich:2007bf}. Here we will adopt the formulae in \cite{Compere:2008cv} where the covariant theory of asymptotic charges is applied specifically for TMG. For the original application in AdS$_3$ using the Hamiltonian formalism see \cite{Brown:1986nw}, while an alternative background independent definition of charges can be found in \cite{Miskovic:2009kr}.

\vspace{3mm}

The charge differences $Q_\xi[g,\bar g]$ between the reference solution $\bar g$ and the solution of interest $g$ are by
\begin{equation}
Q_\xi\,[g ;\bar g] = \int_{\bar g}^g \int_S\, k^{\mu\nu}_\xi[\delta g ; g] \eps_{\mu\nu\rho}\,dx^\rho
\end{equation}
where the first integration is performed in the phase space of solutions, $S$ is a circle at asymptotic infinity, the two-form $k^{\mu\nu}[\delta g,g]\eps_{\mu\nu\rho}dx^\rho$ is a one-form defined in the linearized theory by
\begin{eqnarray}
(16 \pi G)\,k^{\mu\nu}_{\xi}[\delta g;g] &=& (16 \pi G)\,k^{\mu\nu}_{Ein,\xi_{tot}}[\delta g;g]  -\frac{1}{\mu}\xi_\lambda \left( 2 \eps^{\mu\nu\rho}\delta (G^\lambda_{\;\rho}) -  \eps^{\mu\nu\lambda}\delta G )\right) \nonumber \\
 && -\frac{1}{\mu} \eps^{\mu\nu\rho} \left( \xi_\rho h^{\lambda\sigma}G_{\sigma\lambda} +\frac 1 2 h(\xi_\sigma G^\sigma_{\;\rho}+\frac 1 2 \xi_\rho R) \right)\nonumber\\&&\nonumber\\&&+(16 \pi G)\,E^{\mu\nu}[\delta g ;\cL_\xi g],\label{chargetot}
\end{eqnarray}
$\xi_{tot}^{\nu} = \xi^\nu + \frac 1 {2\mu} \eps^{\nu \rho\sigma}D_\rho \xi_\sigma$, the two-form $k^{\mu\nu}_{Ein,\xi}\,[\delta g;g]$ is the Iyer-Wald expression \cite{Iyer:1994ys} for general relativity
\be
(16 \pi G)\sqrt{-g}\,k^{\mu\nu}_{Ein,\xi}\,[\delta g;g] = \sqrt{-g}\,\xi^\mu(D_\lambda h^{\lambda \nu}-D^\lambda h) -\delta(\sqrt{-g}D^\mu \xi^\nu) \,-\, (\mu \leftrightarrow \nu)
\ee
where the variation acts only on $g$, and $E^{\mu\nu}[\delta g ;\cL_\xi \,g]$ is an additional contribution linear in the Killing equation and its derivatives
\begin{eqnarray}
(16 \pi G)E^{\mu\nu}[\delta g ;\cL_\xi\, g] &=& \frac 1 2 \left( h^\nu_{\;\,\lambda}\cL_\xi g^{\lambda \mu} - h^\mu_{\;\,\lambda}\cL_\xi g^{\lambda \nu}\right) +\frac{1}{4\mu}\eps^{\mu\nu\rho}\left(
D_\lambda \cL_\xi g^\lambda_{\;\;\rho}- D_\rho \cL_\xi g_\lambda^{\;\;\lambda}
  \right) h.\nonumber\\\label{defE}
\end{eqnarray}
The charges do not depend on the path chosen in the integration if the integrability condition $ \delta\,\oint_S k^{\mu\nu}_\xi[\delta g ; g] \eps_{\mu\nu\rho}\,dx^\rho = 0$ holds. The crucial property of these charges is that they represent the Lie algebra of asymptotic symmetries via a covariant Poisson bracket up to central charges
\begin{equation}
\{Q_\xi\,[g ;\bar g],Q_{\xi\,^\prime}[g ;\bar g]\} = Q_{[\xi,\,\xi\,^\prime]}[g ;\bar g] + K_{\xi,\,\xi\,^\prime}[\bar g].
\label{eq:repr}
\end{equation}
Here $[\xi,\xi\,^\prime]$ is the Lie bracket and $K_{\xi,\xi\,^\prime}[\bar g] \equiv \int_S k^{\mu\nu}_{\xi}[\cL_{\xi\,^\prime} \bar g;\bar g]$. More precisely this result holds modulo a technical assumption (see (4.3) of \cite{Barnich:2007bf}) which will be checked for the case at hand.

\subsection{Central charge}

Let us denote the charge differences between self-dual warped black
hole metric $g$ and the vacuum ${\bar g}\, (x_+ = x_- = 0,\, \alpha
= 1)$ by $Q_n \equiv Q_{\xi_n}[g;\bar g]$ and $E_R \equiv
Q_{\partial_\tau}[g;\bar g]$. One can check that $Q_n$ is finite
under the boundary condition (\ref{BC}), while the finiteness of
$E_R$ requires an additional boundary condition (\ref{suppBC}). In
order that the asymptotic symmetries form an algebra, additional
conditions on $E^{\mu\nu}$ were suggested in \cite{Barnich:2007bf}
and a simpler sufficient condition was proposed in
\cite{Compere:2007in}, requiring that $\oint_S E^{\mu\nu}[\delta g ;
\cL_\xi \,g]  = 0$. For the generators $Q_n$, using the definition
of $E^{\mu\nu}$ (\ref{defE}) one can check that this condition is
satisfied in the phase space (\ref{BH}). The term $E^{\mu\nu}$
simply does not contribute to any charge. Therefore, the charges
form a representation of the asymptotic symmetry algebra. The
central charge $K_{\xi,\xi\,^\prime}[\bar g] \equiv \int_S
k^{\mu\nu}_{\xi}[\cL_{\xi\,^\prime} \bar g;\bar g]$ could be
calculated by implementing the formula (\ref{chargetot}) in a
Mathematica code. We find the following centrally extended Virasoro
algebra
\begin{eqnarray}
i \{Q_m,Q_n\} &=& (m-n)Q_{m+n}+ \frac{c_L}{12}m(m^2- 2 )\delta_{m+n,0}\,,
\end{eqnarray}
where
\bea
c_L\,=\,\frac{4\nu\ell}{G(\nu^2 + 3)}\,
\eea
is the Virasoro central charge. This is exactly the value of the left central charge conjectured in \cite{Andy08} for spacelike warped AdS$_3$.
Here we have derived it without a Sugawara-type procedure from a current algebra as in \cite{Blagojevic:2009ek}. Since the other Virasoro
sector vanishes we simply set \,$c_R = 0$\,.

\vspace{3mm}

We verify that the entropy of the self-dual warped black hole can be reproduced from the Cardy formula describing the density of states of the dual CFT
\bea
S =\frac { 2\pi\alpha\nu\ell } { 3G(\nu^2 + 3) } = \frac{\pi^2\ell}{3} \frac{4\nu\ell}{G(\nu^2 + 3)} \frac{\alpha}{2\pi\ell}
\equiv\frac{\pi^2\ell}{3} c_L T_L\,,
\eea
which provides strong evidence in favour of the warped AdS/CFT correspondence.

\vspace{3mm}

The boundary conditions \eqref{BC} can be relaxed to $h_{\tau x}=O(1/x)$, in which case we find the asymptotic symmetries are augmented with the vectors
\begin{equation}\label{eq:newAKV}
 \tilde\xi=f(\tau)\,\partial_\tau - f'(\tau)\,x\,\partial_x~.
\end{equation}
The charges associated to the $\tilde\xi$ are finite provided the condition \eqref{suppBC} holds, while the Lie derivative of $\tilde\xi$ on the class of self-dual black holes \eqref{BH} preserves that condition. With
\begin{equation}
 f(\tau)=\left( \tanh\tau+\frac{i}{\cosh\tau}\right)^n \cosh\tau~,n\in\mathbb{Z}~,
\end{equation}
we obtain a second Virasoro that commutes with the first, gives zero charge for the self-dual black holes, and a zero central extension $c_R=0$. Whether or not we include these, the results of this section remain unchanged.

\section{Scalar perturbation}

In order to check the warped AdS/CFT correspondence further, we calculate the real-time correlator of scalar perturbations in this section.
We will show that the results from gravity agree precisely with predictions from the dual CFT description.

\vspace{3mm}

Consider a scalar field $\Phi$ with mass $m$ in the background (\ref{BH}), expanding in modes as
\be\label{ansatz}
\Phi=e^{-i\omega \tau+ik\phi}\cR(x)\,,
\ee
the radial wave function $\cR(x)$ satisfies the equation
\be
\frac{d}{dx}\left((x-x_+)(x-x_-)\frac{d}{dx}\right)\cR(x)-\left(
\frac{\nu^2 + 3}{4\nu^2}\frac{k^2}{\alpha^2} + \frac{\ell^2}{\nu^2 + 3}m^2 - \frac{\left(\omega+\frac{k}{\alpha }(x-\frac{x_+ + x_-}{2})\right)^2}{(x-x_+)(x-x_-)}\right)\cR(x)=0\,.
\ee
We would like to calculate the retarded Green's function, for which purpose the ingoing boundary condition at the horizon is chosen.
Then the solution is
\bea
\cR(x)&=&N\left(\frac{x-x_+}{x-x_-}\right)^{-\frac{i}{2}\left(\frac{k}{\alpha}+\frac{2\omega}{x_+ - x_-}\right)}
\left(\frac{x_+ - x_-}{x-x_-}\right)^{\frac{1}{2}-\beta}\nn\\
&&\quad \cdot ~F\left(\frac{1}{2}-\beta-i\frac{k}{\alpha},\, \frac{1}{2}-\beta-i \frac{2\omega}{x_+ - x_- },\,
1-i\left(\frac{k}{\alpha}+\frac{2\omega}{x_+ - x_-}\right);\, \frac{x-x_+}{x-x_-}\right), \quad
\eea
where $N$ is an arbitrary constant and
\be
\beta^2=\,\frac{1}{4}-\frac{3(\nu^2 -1)}{4\nu^2}\frac{k^2}{\alpha^2} + \frac{\ell^2 }{\nu^2 + 3}m^2.
\ee
At asymptotical infinity, the radial eigenfunction has the behavior
\be\label{asymsol}
\cR(x)\sim Ax^{-\frac{1}{2}-\beta}+Bx^{-\frac{1}{2}+\beta}
\ee
where
\bea
A&=&N\,(x_+ - x_-)^{\frac{1}{2}+\beta}\,\frac{\Gamma\left(-2\beta\right)\Gamma\left(1-i\left(\frac{k}{\alpha}+\frac{2\omega}{x_+ - x_-}\right)\right)}
{\Gamma\left(\frac{1}{2}-\beta-i\frac{k}{\alpha}\right)\Gamma\left(\frac{1}{2}-\beta-i \frac{2\omega}{x_+ - x_-}\right)}\,, \\
 B&=& A\,(\,\beta\to -\beta\,)\,.
 \eea
The real-time retarded Green's function could be computed in terms of the boundary values of the bulk fields using the prescription proposed
 in \cite{Son:2002sd} and later recast in \cite{Iqbal:2008by,Iqbal:2009fd}. This prescription works well not just for asymptotically AdS
metrics, but also for the warped AdS/CFT correspondence\cite{Chen:2009cg} as well as for the Kerr/CFT correspondence\cite{Chen:2010ni,Chen:2010xu}.
However, for the latter cases, a careful study of the holographic renormalization is needed as in the usual AdS/CFT correspondence
\cite{Skenderis:2008dg}. For a scalar field with the asymptotic behavior (\ref{asymsol}), consider a real $\beta>0$ without loss of
generality, the retarded correlator is given by
 \be \label{GR}
 G_R \sim  \frac{A}{B}
  \,=\,(x_+ - x_-)^{2\b}\, \frac{\G(-2\b)}{\G(2\b)}\,\frac{\G\left(\frac{1}{2}+\b-i \frac{k}{\alpha}\right)
\,\G\left(\frac{1}{2}+\b-i \frac{2\omega}{x_+ - x_-}\right)}
  {\G\left(\frac{1}{2}-\b-i \frac{k}{\alpha}\right)\,\G\left(\frac{1}{2}-\b-i \frac{2\omega}{x_+ - x_-}\right)}\,.
 \ee

\vspace{3mm}

On the other hand, throwing a scalar $\Phi$ at the black hole is dual to exciting the CFT by acting with an operator $\cO_\Phi$\,.
For an operator of dimensions $(h_L , h_R)$ at temperature $(T_L, T_R)$,
the momentum-space Euclidean Green's function is determined by conformal invariance and takes the form \cite{Maldacena:1997ih}
\bea \label{GE}
G_E(\o_{L,E}, \o_{R,E})  \sim T_L^{2 h_L-1}  T_R^{2 h_R-1}
e^{i \frac{\o_{L,E}}{2T_L}} e^{i \frac{\o_{L,E}}{2T_R}}
\G(h_L + \frac{\o_{L,E}}{2 \pi T_L})\G(h_L - \frac{\o_{L,E}}{2 \pi T_L})
\nn\\
\cdot\,\G(h_R + \frac{\o_{R,E}}{2 \pi T_R})\G(h_R - \frac{\o_{R,E}}{2 \pi T_R})\,,
\eea
where
 $\o_{L,E},\, \o_{R,E}$ are the Euclidean frequencies, which are related to the Minkowskian ones by $\o_{L,E} = i \o_L\,, ~ \o_{R,E} = i \o_R\,.$
At finite temperature, $\o_{L,E}$ and $\o_{R,E}$ take discrete values of the Matsubara frequencies
$\o_{L,E} =  2 \pi m_L T_L\,, ~ \o_{R,E} =  2 \pi m_R T_R\,,$
where $m_L,\, m_R$ are integers for bosonic modes. The Euclidean correlator $G_E(\o_{L,E}, \o_{R,E})$ is related to the value of the retarded correlator $G_R(\o_{L}, \o_{R})$ by
\be \label{GER}
G_E(\o_{L,E}, \o_{R,E})=G_R(i\o_{L,E}, i\o_{R,E})\, , \quad \o_{L,E} , \o_{R,E} >0.
\ee
Comparing the arguments of the Gamma functions among (\ref{GR}) and (\ref{GER}), one finds precise agreement under the following identification
\be
 h_L=h_R=\frac{1}{2}+\b,~~~{\omega}_L=k/\ell, ~~~{\omega}_R=\omega/\ell, ~~~T_L=\frac{\alpha}{2\pi\ell},
 ~~~T_R = \frac{x_+ - x_-}{4\pi\ell},
\ee
up to an normalization factor. Note that the conformal dimension $h_{L,R}$ above is the same as the ones discussed in \cite{ChenXu2}. As the asymptotic geometry of self-dual warped solutions is the same as the one of global warped AdS$_3$, we do not need to make any extra identification of quantum numbers to find the agreements.

\vspace{3mm}

This argument still holds for imaginary $\beta$, for which the complex conformal weight indicates an instability of the
AdS spacetime due to pair production, similar to the situation of the Kerr/CFT correspondence \cite{Bredberg:2009pv}.

\vspace{3mm}

From the real-time correlator, we can read the greybody factor from its imaginary part, which is in perfect agreement with the CFT prediction as expected. The quasi-normal modes could be read from the poles in the retarded Green's function. These poles are characterized by
\be
k=-i(2\pi T_L l)(n_1+h_L), \hs{5ex} \o=-i(2\pi T_R l)(n_2+h_R),
\ee
where $n_{1,2}$ are non-negative integers. Strictly speaking, only the ones involving the right temperature could be called quasi-normal modes, whose frequencies are vanishing in the extremal limit.

%\vspace{3mm}

%From the retarded correlator (\ref{GR}), we could also read the cross section directly
%\be
%\s ={\rm Im} (G_R) =\frac{(x_+ - x_-)^{2\b}}{2\b \pi
%(\G(2\b))^2}\sinh(\pi(\frac{k}{\alpha}+\frac{2\omega}{x_+ - x_-}))|\G(\frac{1}{2}+\b-i
%\frac{k}{\alpha})\G(\frac{1}{2}+\b-i \frac{2\omega}{x_+ - x_-})|^2.
%\ee

\section{Conclusions and discussions}

In this paper we studied the self-dual warped black hole solutions in topological massive gravity.
These are the TMG analogs of the near-NHEK metric and are asymptotic to the spacelike warped AdS$_3$. We showed that there exist
consistent boundary conditions, for which the asymptotic symmetry generators form one sector of  Virasoro algebra
with central charge $c_L = \frac{4\nu\ell}{G(\nu^2+3)}$. This implies that the black hole quantum states can be identified
with those of a chiral(left) half of a two-dimensional conformal field theory. Our investigation suggests that the quantum topological massive gravity asymptotic  to the same spacelike warped AdS$_3$ in different consistent ways  may be dual to different 2D CFTs.

\vspace{3mm}

The extremal self-dual warped black holes,
which have been exhibited in \cite{Andy08} in different coordinates, are discrete quotients of spacelike warped AdS$_3$
along the $U(1)_L$ isometry, the dual CFT of which has left-moving temperature $T_L={\frac{\alpha}{2\pi\ell}}$ and vanishing
right-moving temperature. For the non-extremal self-dual warped black holes, the right-moving modes are excited and the dual CFT obtains
a non-vanishing right-moving temperature $T_R = \frac{x_+ - x_-}{4\pi\ell}$. For both extremal and non-extremal self-dual warped
 black holes, the entropy is the same and can be reproduced by Cardy's formula albeit using only a left-moving sector. This gives strong
support to the warped AdS/CFT correspondence, and is further corroborated by the perfect agreement of the retarded Green's function from both the gravity computation and the CFT prediction.

\vspace{3mm}

Another interesting feature of the self-dual warped black holes is that they have the same asymptotic geometry as the one of the global warped AdS$_3$. While for the spacelike warped AdS$_3$ black hole, the asymptotic geometry could only be related to the one of global warped spacetime by a local coordinate transformation, but is different globally\cite{Anninos:2009zi}. In other words, the global warped AdS$_3$ is not the ground state of the warped AdS$_3$ black holes. In our case, we can take the global warped AdS$_3$ as the ground state after periodically identifying $u$.

\vspace{3mm}

The warped AdS/CFT correspondence proposed here is a chiral one, very different from the one suggested in \cite{Andy08}. The boundary conditions we proposed are different from those suggested in \cite{Compere:2008cv,Compere:2009zj}. It is remarkable that we obtain the left central charge conjectured in \cite{Andy08}
naturally through a centrally extended Virasoro algebra which is an enhancement of the $U(1)_L$ isometry, instead of a Sugawara-type
procedure from a current algebra as in \cite{Blagojevic:2009ek}. Since for non-extremal self-dual warped black holes,
the right-moving temperature doesn't vanish, it is natural to expect
from the Cardy formula that the dual CFT has a right central charge $c_R = 0$. For the boundary conditions (\ref{BC}) (\ref{suppBC}),
the right-moving Virasoro sector vanishes. We leave open
the question of whether another set of boundary conditions, such as the one leading to \eqref{eq:newAKV}, admits a consistent Virasoro sector
with central charge $c_R = 0$\,\cite{Compere:2009zj}.

\vspace{3mm}

In the Kerr/CFT correspondence, the left central charge could be obtained from the asymptotic symmetry of NHEK. The right central charge could be read from the study of AdS$_2$ gravity after reduction\cite{castro-larsen}. It would be interesting to study the relation between TMG and AdS$_2$ gravity and understand the absence of right central charge\cite{Chen2010}.

\vspace{3mm}

We computed the scalar real-time correlators in the self-dual warped black hole and found perfect agreement with the CFT prediction. It would also be interesting to check the warped AdS/CFT correspondence further by computing the
real-time correlators for vector, spinor as well as gravitational perturbations.

\section*{Acknowledgments}

We are very grateful to G. Moutsopoulos for his contributions and
comments on this project.
% BC would like to thank KIAS for hospitality during his visit.
The work was partially supported by NSFC Grant No.10775002,10975005 and NKBRPC (No. 2006CB805905).

¡¡

\ed
\begin{thebibliography}{99}

%\cite{Deser:1981wh}
\bibitem{Deser:1981wh}
  S.~Deser, R.~Jackiw and S.~Templeton,
  ``Topologically massive gauge theories,''
  Annals Phys.\  {\bf 140}, 372 (1982)
  %[Erratum-ibid.\  {\bf 185}, 406.1988\ APNYA,281,409 (1988\ APNYA,281,409-449.2000)].
  %%CITATION = APNYA,281,409;%%

  %\cite{Deser:1982vy}
\bibitem{Deser:1982vy}
  S.~Deser, R.~Jackiw and S.~Templeton,
  ``Three-Dimensional Massive Gauge Theories,''
  Phys.\ Rev.\ Lett.\  {\bf 48}, 975 (1982).
  %%CITATION = PRLTA,48,975;%%

%\cite{Li:2008dq}
\bibitem{Li:2008dq}
  W.~Li, W.~Song and A.~Strominger,
  ``Chiral Gravity in Three Dimensions,''
  JHEP {\bf 0804}, 082 (2008)
  [arXiv:0801.4566 [hep-th]].
  %%CITATION = JHEPA,0804,082;%%


\bibitem{Andy08}
  D.~Anninos, W.~Li, M.~Padi, W.~Song and A.~Strominger,
  ``Warped AdS$_3$ Black Holes,''
  JHEP {\bf 0903}, 130 (2009)
  [arXiv:0807.3040 [hep-th]].
  %%CITATION = JHEPA,0903,130;%%

%\cite{Compere:2008cv}
\bibitem{Compere:2008cv}
  G.~Compere and S.~Detournay,
  ``Semi-classical central charge in topologically massive gravity,''
  Class.\ Quant.\ Grav.\  {\bf 26} (2009) 012001
  %[Erratum-ibid.\  {\bf 26} (2009) 139801]
  [arXiv:0808.1911 [hep-th]].
  %%CITATION = CQGRD,26,012001;%%

%\cite{Compere:2009zj}
\bibitem{Compere:2009zj}
  G.~Compere and S.~Detournay,
  ``Boundary conditions for spacelike and timelike warped AdS$_3$ spaces in
  topologically massive gravity,''
  JHEP {\bf 0908} (2009) 092
  [arXiv:0906.1243 [hep-th]].
  %%CITATION = JHEPA,0908,092;%%

%\cite{Blagojevic:2009ek}
\bibitem{Blagojevic:2009ek}
  M.~Blagojevic and B.~Cvetkovic,
  ``Asymptotic structure of topologically massive gravity in spacelike
  stretched AdS sector,''
  JHEP {\bf 0909} (2009) 006
  [arXiv:0907.0950 [gr-qc]].
  %%CITATION = JHEPA,0909,006;%%

%\cite{Chen:2009rf}
\bibitem{ChenXu09}
  B.~Chen and Z.-b.~Xu,
  ``Quasinormal modes of warped $AdS_3$ black holes and AdS/CFT
  correspondence,''
   [arXiv:0901.3588 [hep-th]].
  %%CITATION = ARXIV:0901.3588;%%

%\cite{Chen:2009hg}
\bibitem{ChenXu2}
  B.~Chen and Z.-b.~Xu,
  ``Quasi-normal modes of warped black holes and warped AdS/CFT
  correspondence,''
  JHEP {\bf 0911}, 091 (2009)
  [arXiv:0908.0057 [hep-th]].
  %%CITATION = JHEPA,0911,091;%%

%\cite{Chen:2009cg}
\bibitem{Chen:2009cg}
  B.~Chen, B.~Ning and Z.-b.~Xu,
  ``Real-time correlators in warped AdS/CFT correspondence,''
  JHEP {\bf 1002}, 031 (2010)
  [arXiv:0911.0167 [hep-th]].
  %%CITATION = JHEPA,1002,031;%%

%\cite{Bredberg:2009pv}
\bibitem{Bredberg:2009pv}
  I.~Bredberg, T.~Hartman, W.~Song and A.~Strominger,
  ``Black Hole Superradiance From Kerr/CFT,''
  JHEP {\bf 1004}, 019 (2010)
  [arXiv:0907.3477 [hep-th]].
  %%CITATION = JHEPA,1004,019;%%

%\cite{Cardy:1986ie}
\bibitem{Cardy:1986ie}
  J.~L.~Cardy,
  ``Operator Content Of Two-Dimensional Conformally Invariant Theories,''
  Nucl.\ Phys.\  B {\bf 270}, 186 (1986).
  %%CITATION = NUPHA,B270,186;%%

%\cite{Guica:2008mu}
\bibitem{Guica:2008mu}
  M.~Guica, T.~Hartman, W.~Song and A.~Strominger,
  ``The Kerr/CFT Correspondence,''
  Phys.\ Rev.\  D {\bf 80}, 124008 (2009)
  [arXiv:0809.4266 [hep-th]].
  %%CITATION = PHRVA,D80,124008;%%

%\cite{Spradlin:1999bn}
\bibitem{Spradlin:1999bn}
  M.~Spradlin and A.~Strominger,
  ``Vacuum states for AdS(2) black holes,''
  JHEP {\bf 9911}, 021 (1999)
  [arXiv:hep-th/9904143].
  %%CITATION = JHEPA,9911,021;%%

%\cite{Coussaert:1994tu}
\bibitem{Coussaert:1994tu}
  O.~Coussaert and M.~Henneaux,
  ``Self-dual solutions of 2+1 Einstein gravity with a negative  cosmological
  constant,''
  [arXiv:hep-th/9407181].
  %%CITATION = HEP-TH/9407181;%%

  %\cite{Bouchareb:2007yx}
\bibitem{Bouchareb:2007yx}
  A.~Bouchareb and G.~Clement,
  ``Black hole mass and angular momentum in topologically massive gravity,''
  Class.\ Quant.\ Grav.\  {\bf 24}, 5581 (2007)
  [arXiv:0706.0263 [gr-qc]].
  %%CITATION = CQGRD,24,5581;%%

    %\cite{Abbott:1982jh}
\bibitem{Abbott:1982jh}
  L.~F.~Abbott and S.~Deser,
  ``Charge Definition In Nonabelian Gauge Theories,''
  Phys.\ Lett.\  B {\bf 116}, 259 (1982).
  %%CITATION = PHLTA,B116,259;%%

  %\cite{Deser:2002rt}
\bibitem{Deser:2002rt}
  S.~Deser and B.~Tekin,
  ``Gravitational energy in quadratic curvature gravities,''
  Phys.\ Rev.\ Lett.\  {\bf 89}, 101101 (2002)
  [arXiv:hep-th/0205318].
  %%CITATION = PRLTA,89,101101;%%

  %\cite{Deser:2002jk}
\bibitem{Deser:2002jk}
  S.~Deser and B.~Tekin,
  ``Energy in generic higher curvature gravity theories,''
  Phys.\ Rev.\  D {\bf 67}, 084009 (2003)
  [arXiv:hep-th/0212292].
  %%CITATION = PHRVA,D67,084009;%%

  %\cite{Deser:2003vh}
\bibitem{Deser:2003vh}
  S.~Deser and B.~Tekin,
  ``Energy in topologically massive gravity,''
  Class.\ Quant.\ Grav.\  {\bf 20}, L259 (2003)
  [arXiv:gr-qc/0307073].
  %%CITATION = CQGRD,20,L259;%%

%\cite{Solodukhin:2005ah}
\bibitem{Solodukhin:2005ah}
  S.~N.~Solodukhin,
  ``Holography with gravitational Chern-Simons,''
  Phys.\ Rev.\  D {\bf 74}, 024015 (2006)
  [arXiv:hep-th/0509148].
  %%CITATION = PHRVA,D74,024015;%%

%\cite{Tachikawa:2006sz}
\bibitem{Tachikawa:2006sz}
  Y.~Tachikawa,
  ``Black hole entropy in the presence of Chern-Simons terms,''
  Class.\ Quant.\ Grav.\  {\bf 24}, 737 (2007)
  [arXiv:hep-th/0611141].
  %%CITATION = CQGRD,24,737;%%

%\cite{Frolov:1989jh}
\bibitem{Frolov:1989jh}
  V.~P.~Frolov and K.~S.~Thorne,
  ``Renormalized Stress-Energy Tensor Near the Horizon of a Slowly Evolving, Rotating Black Hole''
  Phys.\ Rev.\  D {\bf 39}, 2125 (1989).
  %%CITATION = PHRVA,D39,2125;%%

%\cite{Ottewill:2000qh}
\bibitem{Ottewill:2000qh}
  A.~C.~Ottewill and E.~Winstanley,
  ``The renormalized stress tensor in Kerr space-time: General results,''
  Phys.\ Rev.\  D {\bf 62}, 084018 (2000)
  [arXiv:gr-qc/0004022].
  %%CITATION = PHRVA,D62,084018;%%
%\cite{Duffy:2005mz}

\bibitem{Duffy:2005mz}
  G.~Duffy and A.~C.~Ottewill,
  ``The renormalized stress tensor in Kerr space-time: Numerical results  for
  the Hartle-Hawking vacuum,''
  Phys.\ Rev.\  D {\bf 77}, 024007 (2008)
  [arXiv:gr-qc/0507116].
  %%CITATION = PHRVA,D77,024007;%%

%\cite{Maldacena:1998uz}
\bibitem{Maldacena:1998uz}
  J.~M.~Maldacena, J.~Michelson and A.~Strominger,
  ``Anti-de Sitter fragmentation,''
  JHEP {\bf 9902}, 011 (1999)
  [arXiv:hep-th/9812073].
  %%CITATION = JHEPA,9902,011;%%

%\cite{Clement:2004yr}
\bibitem{Clement:2004yr}
  G.~Clement and C.~Leygnac,
  ``Non-asymptotically flat, non-AdS dilaton black holes,''
  Phys.\ Rev.\  D {\bf 70}, 084018 (2004)
  [arXiv:gr-qc/0405034].
  %%CITATION = PHRVA,D70,084018;%%

%\cite{Moussa:2008sj}
\bibitem{Moussa:2008sj}
  K.~A.~Moussa, G.~Clement, H.~Guennoune and C.~Leygnac,
  ``Three-dimensional Chern-Simons black holes,''
  Phys.\ Rev.\  D {\bf 78}, 064065 (2008)
  [arXiv:0807.4241 [gr-qc]].
  %%CITATION = PHRVA,D78,064065;%%

%\cite{Banados:1992gq}
\bibitem{Banados:1992gq}
  M.~Banados, M.~Henneaux, C.~Teitelboim and J.~Zanelli,
  ``Geometry of the (2+1) black hole,''
  Phys.\ Rev.\  D {\bf 48}, 1506 (1993)
  [arXiv:gr-qc/9302012].
  %%CITATION = PHRVA,D48,1506;%%

\bibitem{Maldacena:1998bw}
  J.~M.~Maldacena and A.~Strominger,
  ``AdS(3) black holes and a stringy exclusion principle,''
  JHEP {\bf 9812}, 005 (1998)
  [arXiv:hep-th/9804085].
  %%CITATION = JHEPA,9812,005;%%

%\cite{Barnich:2001jy}
\bibitem{Barnich:2001jy}
  G.~Barnich and F.~Brandt,
  ``Covariant theory of asymptotic symmetries, conservation laws and  central
  charges,''
  Nucl.\ Phys.\  B {\bf 633}, 3 (2002)
  [arXiv:hep-th/0111246].
  %%CITATION = NUPHA,B633,3;%%

%\cite{Barnich:2003xg}
\bibitem{Barnich:2003xg}
  G.~Barnich,
  ``Boundary charges in gauge theories: Using Stokes theorem in the bulk,''
  Class.\ Quant.\ Grav.\  {\bf 20}, 3685 (2003)
  [arXiv:hep-th/0301039].
  %%CITATION = CQGRD,20,3685;%%

%\cite{Barnich:2007bf}
\bibitem{Barnich:2007bf}
  G.~Barnich and G.~Compere,
  ``Surface charge algebra in gauge theories and thermodynamic integrability,''
  J.\ Math.\ Phys.\  {\bf 49}, 042901 (2008)
  [arXiv:0708.2378 [gr-qc]].
  %%CITATION = JMAPA,49,042901;%%

%\cite{Brown:1986nw}
\bibitem{Brown:1986nw}
  J.~D.~Brown and M.~Henneaux,
  ``Central Charges in the Canonical Realization of Asymptotic Symmetries: An
  Example from Three-Dimensional Gravity,''
  Commun.\ Math.\ Phys.\  {\bf 104}, 207 (1986).
  %%CITATION = CMPHA,104,207;%%

%\cite{Miskovic:2009kr}
\bibitem{Miskovic:2009kr}
  O.~Miskovic and R.~Olea,
  ``Background-independent charges in Topologically Massive Gravity,''
  JHEP {\bf 0912}, 046 (2009)
  [arXiv:0909.2275 [hep-th]].
  %%CITATION = JHEPA,0912,046;%%

%\cite{Iyer:1994ys}
\bibitem{Iyer:1994ys}
  V.~Iyer and R.~M.~Wald,
  ``Some properties of Noether charge and a proposal for dynamical black hole
  entropy,''
  Phys.\ Rev.\  D {\bf 50}, 846 (1994)
  [arXiv:gr-qc/9403028].
  %%CITATION = PHRVA,D50,846;%%

%\cite{Compere:2007in}
\bibitem{Compere:2007in}
  G.~Compere and S.~Detournay,
  ``Centrally extended symmetry algebra of asymptotically Goedel spacetimes,''
  JHEP {\bf 0703}, 098 (2007)
  [arXiv:hep-th/0701039].
  %%CITATION = JHEPA,0703,098;%%

%\cite{Son:2002sd}
\bibitem{Son:2002sd}
  D.~T.~Son and A.~O.~Starinets,
  ``Minkowski-space correlators in AdS/CFT correspondence: Recipe and
  applications,''
  JHEP {\bf 0209}, 042 (2002)
  [arXiv:hep-th/0205051].
  %%CITATION = JHEPA,0209,042;%%

%\cite{Iqbal:2008by}
\bibitem{Iqbal:2008by}
  N.~Iqbal and H.~Liu,
  ``Universality of the hydrodynamic limit in AdS/CFT and the membrane
  paradigm,''
  Phys.\ Rev.\  D {\bf 79}, 025023 (2009)
  [arXiv:0809.3808 [hep-th]].
  %%CITATION = PHRVA,D79,025023;%%

%\cite{Iqbal:2009fd}
\bibitem{Iqbal:2009fd}
  N.~Iqbal and H.~Liu,
  ``Real-time response in AdS/CFT with application to spinors,''
  Fortsch.\ Phys.\  {\bf 57}, 367 (2009)
  [arXiv:0903.2596 [hep-th]].
  %%CITATION = FPYKA,57,367;%%

%\cite{Chen:2010ni}
\bibitem{Chen:2010ni}
  B.~Chen and C.~S.~Chu,
  ``Real-time correlators in Kerr/CFT correspondence,''
  JHEP {\bf 1005}, 004 (2010)
  [arXiv:1001.3208 [hep-th]].
  %%CITATION = JHEPA,1005,004;%%

%\cite{Chen:2010xu}
\bibitem{Chen:2010xu}
  B.~Chen and J.~Long,
  ``Real-time Correlators and Hidden Conformal Symmetry in Kerr/CFT
  Correspondence,''
  [arXiv:1004.5039 [hep-th]].
  %%CITATION = ARXIV:1004.5039;%%

%\cite{Skenderis:2008dg}
\bibitem{Skenderis:2008dg}
  K.~Skenderis and B.~C.~van Rees,
  ``Real-time gauge/gravity duality: Prescription, Renormalization and
  Examples,''
  JHEP {\bf 0905}, 085 (2009)
  [arXiv:0812.2909 [hep-th]].
  %%CITATION = JHEPA,0905,085;%%

%\cite{Maldacena:1997ih}
\bibitem{Maldacena:1997ih}
  J.~M.~Maldacena and A.~Strominger,
  ``Universal low-energy dynamics for rotating black holes,''
  Phys.\ Rev.\  D {\bf 56}, 4975 (1997)
  [arXiv:hep-th/9702015].
  %%CITATION = PHRVA,D56,4975;%%

%\cite{Anninos:2009zi}
\bibitem{Anninos:2009zi}
  D.~Anninos, M.~Esole and M.~Guica,
  ``Stability of warped AdS3 vacua of topologically massive gravity,''
  JHEP {\bf 0910}, 083 (2009)
  [arXiv:0905.2612 [hep-th]].
  %%CITATION = JHEPA,0910,083;%%


\bibitem{castro-larsen}
  A.~Castro and F.~Larsen,
  ``Near Extremal Kerr Entropy from AdS$_2$ Quantum Gravity,''
  JHEP {\bf 0912}, 037 (2009)
  [arXiv:0908.1121 [hep-th]].
  %%CITATION = JHEPA,0912,037;%%


\bibitem{Chen2010}
  Work in Progress.


\end{thebibliography}
